\documentclass[12pt]{amsart}

\usepackage{epsfig}

\begin{document}

\title[]{\bf Perturbative Evaluation of the 
	Zero-Point function for Self-Interacting Scalar Field
	on a Manifold with Boundary}

\author[]{George Tsoupros \\
       {\em School of Physics}\\
       {\em The University of New South Wales}\\
       {\em NSW 2052}\\
       {\em Australia}
}
\thanks{present e-mail address: gts@phys.unsw.edu.au}

\begin{abstract}

The character of quantum corrections to the gravitational action of a conformally 
invariant field theory for a self-interacting scalar field on a manifold with boundary 
is considered at third loop-order in the perturbative expansion of the zero-point 
function. Diagramatic evaluations and higher loop-order renormalisation can be best 
accomplished on a Riemannian manifold of positive constant curvature accommodating a boundary 
of constant extrinsic curvature. The associated spherical formulation for diagramatic 
evaluations reveals a non-trivial effect which the topology of the manifold has on 
the vacuum processes and which ultimately dissociates the dynamical behaviour of the 
quantised field from its behaviour in the absence of a boundary. The first surface 
divergence is evaluated and the necessity for simultaneous renormalisation of volume 
and surface divergences is shown. 

\end{abstract}

\maketitle


{\bf I. Introduction}\\

The development of Euclidean Quantum Gravity has emphasised the importance of any investigation 
relevant to the dynamical behaviour of quantised matter fields on a manifold with boundary by 
revealing the latter's physical significance in the quantum cosmological context of semiclassical 
tunnelling. Specifically, instanton-related considerations in quantum and inflationary cosmology 
lend particular importance to the issue of radiative contributions to a semiclassical tunnelling 
geometry of positive constant curvature effected by quantised matter. The presence of a boundary 
drastically alters the dynamical behaviour of the quantum field. In addition to the contributions 
which any divergence stemming from vacuum effects at any loop-order receives from the Riemann 
curvature of the background geometry the same divergence receives contributions from the extrinsic 
curvature of the boundary. This effect which has been studied at one-loop order through heat kernel 
techniques \cite{AvraEspo} will be shown in this project to be relevant to any loop-order. Its 
occurrence is exemplary of the potential effect which the topology of a manifold has on the short 
distance dynamical behaviour of the quantum field defined on it.

Any evaluation of higher loop-order vacuum effects generated on the dynamical behaviour 
of quantised matter on a manifold with boundary necessitates diagramatic techniques. 
Diagramatic evaluations and higher loop-order renormalisation can be best accomplished 
on a Riemannian manifold of positive constant curvature accommodating a boundary of constant 
extrinsic curvature. The advantage of working on a Euclidean n-dimensional spherical cap $ C_n$ stems 
from its underlying symmetry which allows for the exploitation of the method of images, thereby 
allowing for the exploitation of the emerging spherical symmetry in the evaluation of radiative 
effects. In the physically important case of a massless scalar field conformally coupled to 
the background geometry of $ C_n$ with the Dirichlet condition specified on the boundary $ \partial C_n$ 
such an approach has resulted in an exact expression for the massless propagator \cite{George} \cite{Tsoupros}. 
This, in turn, allows for the development of the diagramatic techniques necessary for the evaluation of 
higher loop-order radiative contributions to the associated effective action for a self-interacting 
conformal scalar field at $ n=4$. A prerequisite for such an evaluation is a spherical formulation of the 
Feynman rules on $ C_n$. Such a formulation has been advanced in the context of the method of images by 
exploiting the association of the dynamical behaviour of the conformal scalar field on $ C_n$ to that of 
the same field on the Euclidean n-sphere $ S_n$ \cite{Tsoupros}. Consequent upon such a formulation was, 
in addition, a confirmation of the results obtained in \cite{George} to the effect that, at any loop-order, 
no independent renormalisation is necessary. The volume-related terms in the semiclassical action receive 
infinite contributions from vacuum effects simultaneously with the boundary-related terms. In this respect, 
the stated spherical formulation confirmed, in addition, the potential for non-trivial radiative generation 
of surface counterterms in the effective action.
  
The assessment of the stated boundary-related and vacuum effects will be pursued, in what follows, by 
exploiting the results obtained in \cite{Tsoupros} in order to evaluate the zero-point function at
third loop-order (second order in the self-coupling) for a self-interacting massless scalar field conformally 
coupled to the background geometry of $ C_4$. Such an evaluation will also constitute the incipient point 
of a renormalisation program to the same order. From the outset, all divergences in the gravitational action 
generated by scalar loops on a four dimensional manifold are, on dimensional grounds, expected to be 
proportional to the cosmological constant $ \Lambda$, to the Ricci scalar $ R$, to $ R^2$, and to the Ricci 
and Riemann tensor-related contractions $ R_{\mu\nu}R^{\mu\nu}$ and $
R_{\alpha\beta\gamma\delta}R^{\alpha\beta\gamma\delta}$ respectively. Suitable linear combinations of 
these terms are also possible \cite{Birrel}. If a boundary is present additional divergent contributions 
involving the extrinsic curvature $ K$ of that boundary are expected \cite{George}. The advantage of evaluating 
such gravitational divergences on a manifold of constant curvature is that the concommitant spherical 
formulation of the theory determines each Green function as an exact function of the constant Ricci scalar 
$ R$ \cite{McKeon Tsoupros}. This result has both formal and perturbative significance and constitutes, for 
that matter, the primary motivation for the entire spherical formulation of euclidean quantum field theory 
\cite{Drummond}, \cite{I.Drummond}, \cite{DrummondShore}, \cite{Shore}, \cite{G.Shore}. Effectively, only 
volume-related counterterms proportional to $ \Lambda$, $ R$, $ R^2$ and $ RK^2$ are allowed on $ C_4$. This 
is the case because, perturbatively, the only source of volume divergences are terms in the relevant 
expansions of the Green functions.

The effect which a boundary in space-time has on the dynamical behaviour of quantised matter can be best 
appreciated through a comparison with the same dynamical behaviour in the absence of a boundary. In the 
present case, such a comparison is facilitated by the fact that both $ C_4$ and $ S_4$ are manifolds of 
positive constant curvature. In what follows it will be shown that the divergences inherent in the zero-point
function evaluated at three-loop level receive a non-trivial 
contribution from the boundary $ \partial C_4$ which, effectively, dissociates the dynamical 
behaviour of the quantised scalar field on $ C_4$ from that on $ S_4$. It is also worth noting 
in this respect, that since divergences in the evaluation of loop effects stem essentially from 
the coincidence of integration points in the relevant Feynman diagrams, the emergence of divergences  
proportional to $ K$ signifies a non-trivial effect which the topology of a bounded manifold has on the 
cut-off scale behaviour of the quantised matter fields. The announced simultaneous renormalisation of 
volume and surface terms will, also, become manifest in the context of the perturbative calculation pursued in 
this project. Moreover, the stated potential of the theory for non-trivial generation of surface counterterms 
in the effective action will be realised in the evaluation of the first surface divergence on a manifold with 
boundary.

{\bf II. The Zero Point Function}\\

The unique conformally invariant scalar action on a n-dimensional Riemannian 
manifold of positive constant embedding radius $ \it{a}$ is \cite{McKeon Tsoupros}
 
\begin{equation}
S[{\Phi}] = \int_C{d\sigma}[\frac{1}{2} \frac{1}{2a^2}
\Phi(L^2- \frac{1}{2}n(n-2) )\Phi - \frac{\lambda}{\Gamma(p+1)}\Phi^p]
\end{equation}
provided that $ p = \frac{2n}{n-2}, n>2$. $ L_{\mu \nu}$ is the generator of 
rotations 

\begin{equation}
L_{\mu \nu} = \eta_{\mu}\frac{\partial}{\partial \eta_{\nu}} -
\eta_{\nu}\frac{\partial}{\partial \eta_{\mu}} 
\end{equation}
on the relevant embedded manifold. In the physically relevant case of $ n=4$ the 
self-coupling is that of $ \Phi^4$. Any n-dimensional Riemannian manifold of positive 
constant embedding radius $ \it{a}$ represents, through the usual rotation in imaginary 
time, the Euclidean version of the associated segment of de Sitter space. On any such 
manifold the Ricci scalar $ R$ admits the constant value \cite{McKeon Tsoupros}

\begin{equation}
R = \frac{n(n-1)}{a^2}
\end{equation}
In conformity with the considerations hitherto outlined the Riemannian manifold relevant 
to (1) will be specified to be that of a spherical n-cap $ \it{C_n}$ considered as a 
manifold of positive constant curvature embedded in a $ (n+1)$-dimensional Euclidean space and 
bounded by a $ (n-1)$-sphere of positive constant extrinsic curvature $ K$ (diverging normals). 
In effect, $ \it{C_n}$ is characterised by spherical $(n-1)$-dimensional sections of 
constant Euclidean time $ \tau$.
The Einstein-Hilbert action, being the gravitational component of the conformally 
invariant (semi)classical action, necessitates an additional term of the form 
$ \int_{\partial C} K{\Phi^2}$ enforced by the presence of the boundary $ \partial C$ 
\cite{George}. As a result, the Einstein-Hilbert action at $ n=4$ assumes the form

\begin{equation}
S_{EH} = -\frac{1}{16\pi G} \int_{C}{d\sigma}(R-2\Lambda)+
\int_{{\partial}C}{d^3}x \sqrt{h}K{\Phi^2} 
\end{equation}
with $ d\sigma = a^nd\Omega_{n+1}$ being the element of surface area of the 
n-sphere embedded in $ n+1$ dimensions and with the three-dimensional boundary 
hypersurface on which the induced metric is $ h_{ij}$ being characterised by 
an extrinsic curvature $ K_{ij}=\frac{1}{2}(\nabla_{i}n_{j}+\nabla_{j}n_{i})$
the trace of which is $ K=h^{ij}K_{ij}>0$. Although the surface term in (4) stems
from a variational demand posed on $ S_{EH}$ it is not purely geometrical. For that matter
it is considered an additional sector to the scalar component of the action expressed in (1). 
The action $ S[\Phi]$ for the specified theory is the sum-total of all sectors in (1) and (4).  
In addition to the cosmological constant and scalar curvature featured in $ S_{EH}$ the 
semiclassical approach, which differentiates between the matter and gravitational degrees of 
freedom by quantising the former on a fixed geometrical background, is known to result in all 
possible volume divergences which are dimensionally consistent in the gravitational action. 
These higher-order corrections are effected by vacuum processes of the interacting field 
\cite{BrowColl}. The immediate issue which the presence of a boundary raises perturbatively is 
the possibility of additional divergent contributions in the bare gravitational action featuring 
the extrinsic curvature $ K$. In fact, such contributions to the two-point function - and, for that 
matter, to the bare scalar action - are expected past one-loop order \cite{George}.   

All Green functions of $ \Phi^4$ theory on $ C_4$ are generated by the Euclidean functional 
integral

\begin{equation}
Z[J] = \int D[{\Phi}]e^{S[{\Phi}] + J\Phi }
\end{equation} 
through functional differentiation at $ J=0$. This generating functional is the 
mathematical expression for the transition amplitude between vacuum states defined 
on $ C_4$ \cite{Birrel}. The disconnected components of this generating functional 
in curved space-times - represented by the zero-point function - invariably have 
non-trivial contributions to that transition amplitude and, for that matter, to the 
effective action. This is, essentially, due to graviton contributions which the 
matter propagators receive from the semiclassical background geometry and which, for 
that matter, render a purely geometric character to any divergences inherent in the 
zero-point function. Although stemming from the dynamical behaviour of the matter 
fields these divergences result in the infinite redefinitions in the gravitational 
component of the semiclassical action.    
  
Pursuant to the spherical formulation relevant to diagramatic evaluations on $ C_n$ 
developed in \cite{Tsoupros} all calculations will be advanced in configuration space.
The first two terms in the perturbative expansion of the zero-point function in
powers of $ \hbar$ (loop expansion) are diagramatically represented by the graphs of
fig.(1). The ``bubble'' diagrams in fig.(1a) and fig.(1b) account for the one-loop 
contribution to the zero-point function of the theory. Their simultaneous presence in any 
curved space-time is expected on the basis of general theoretical considerations 
\cite{Birrel}. They are characterised by the absence of interaction vertices and, on power 
counting grounds, are responsible for the simultaneous one-loop contributions to volume 
and boundary effective Einstein-Hilbert action on any manifold with boundary 
\cite{Solodukhin}. They have been shown to be finite provided that dimensional 
regularisation is used \cite{George}. The latter technique regulates all ultraviolet 
divergences through an analytical extension of the space-time dimensionality to an 
arbitrary value $ n$ \cite{'t Hooft}. 

\begin{figure}[h]
\centering\epsfig{figure=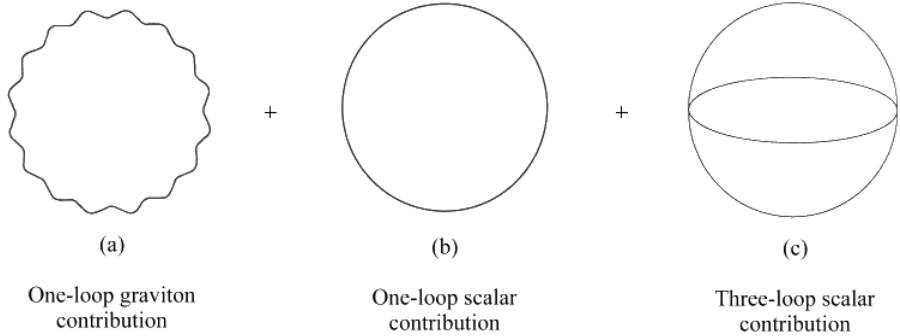, height = 45mm,width=120mm}
\caption{$ O(\hbar^3)$-related contributions to the zero-point function}
\end{figure}

The presence of the self-coupling in the semi-classical scalar action generates additional 
contributions to the zero-point function at two and three loop-level represented by fig.(1d) 
and fig.1(f). They are finite in the context of dimensional regularisation and, for that 
matter, just as trivial as those relevant to fig.(1a) and fig.(1b) 

\begin{figure}[h]
\centering\epsfig{figure=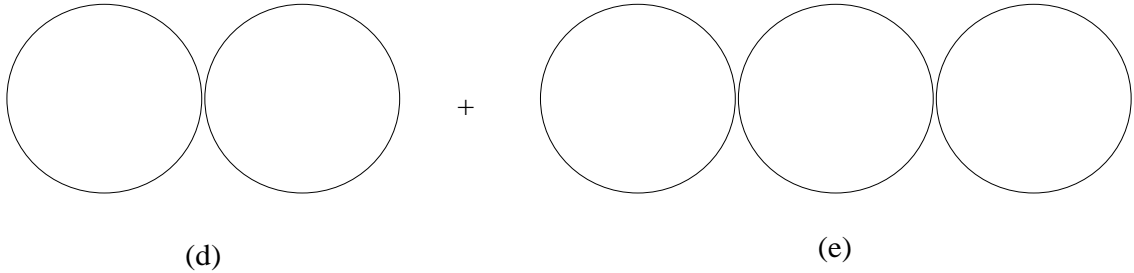, height = 35mm,width=150mm}
\caption{$ O(\hbar^3)$-related finite contributions to the zero-point function}
\end{figure}         
 
The remaining diagram in fig.(1c) represents the first potentially non-trivial 
contribution in the perturbative expansion of the zero-point function. It is of 
order $ \hbar^3$ and features two interaction vertices. In view of the minimal 
subtraction-related expansion 

\begin{equation}
\lambda_0 = \mu^{4-n}[\lambda +
\sum_{\nu=1}^{\infty}\frac{a_{\nu}(\lambda)}{(4-n)^{\nu}}]
\end{equation}
which the bare self-coupling $ \lambda_0$ admits in powers of the renormalised 
self-coupling $ \lambda$ \cite{'t Hooft} the multiplicative factor of $ (-\lambda_0)$ 
dictated by the Feynman rules for each of the two vertices of fig.(1c) amounts, at 
that three-loop order, to $ \lambda^2$ \cite{McKeon Tsoupros}. In effect, the diagram
in question featuring four internal propagators joining the vertices $ \eta$ and 
$ \eta'$ results in the following mathematical expression for the stated 
$ O(\hbar^3)$-term 

\begin{equation}
I_c = \lambda^2 B_n^4 \int d^n{\eta}d^n{\eta'} [D_{c}^{(n)}(\eta,{\eta}') ]^4
\end{equation}
with 

\begin{equation}
B_n =  \frac{\Gamma(\frac{n}{2}-1)}{4\pi^{\frac{n}{2}}}
\end{equation}

The multiplicative factor of $ B_n$ is the symmetry factor associated with the propagator 
$ D_{c}^{(n)}(\eta,{\eta}')$ for the conformal field $ \Phi$ on $ C_n$
which was shown to be \cite{George}, \cite{Tsoupros}

\begin{equation}
D_{c}^{(n)}(\eta,{\eta}') = \frac{\Gamma(\frac{n}{2}-1)}{4\pi^{\frac{n}{2}}}\frac{1}{|{\eta}-{\eta}'|^{n-2}} 
- \frac{\Gamma(\frac{n}{2}-1)}{4\pi^{\frac{n}{2}}}\frac{1}{|\frac{a_{{\eta}'}}{a_B}{\eta}-
\frac{a_B}{a_{{\eta}'}}{\eta}'|^{n-2}}
\end{equation}
with $ a_B$ and $ a_{\eta'}$ being the geodesic distances between the cap's pole and its boundary 
as well as between the former and point $ \eta'$ respectively. Replacing this expression in that 
for $ I_c$ and expanding yields


$$
I_c = \lambda^2 B_n^4 {\int}_C d^n{\eta}d^n{\eta'}\big{[}|{\eta}-{\eta}'|^{4(2-n)} + 
4|{\eta}-{\eta}'|^{3(2-n)}|\frac{a_{{\eta}'}}{a_B}{\eta}-
\frac{a_B}{a_{{\eta}'}}{\eta}'|^{2-n} + $$

$$
6|{\eta}-{\eta}'|^{2(2-n)}|\frac{a_{{\eta}'}}{a_B}{\eta}-
\frac{a_B}{a_{{\eta}'}}{\eta}'|^{2(2-n)} + $$

\begin{equation}
4|{\eta}-{\eta}'|^{2-n}|\frac{a_{{\eta}'}}{a_B}{\eta}-
\frac{a_B}{a_{{\eta}'}}{\eta}'|^{3(2-n)} + 
|\frac{a_{{\eta}'}}{a_B}{\eta}-
\frac{a_B}{a_{{\eta}'}}{\eta}'|^{4(2-n)}\big{]}
\end{equation}

All five terms in this double volume integral will have to be examined separately. Their 
evaluation necessitates the results obtained through a spherical formulation for the 
mathematical expressions of Feynman diagrams on $ C_n$ \cite{Tsoupros}. In the ensuing
calculations use will be made of the expressions

\begin{equation}
[(\eta - \eta')^2]^{\nu} = \sum_{N=0}^{\infty}\sum_{\alpha=0}^{N}
\frac{(2a)^{2\nu+n}{\pi}^{\frac{n}{2}}\Gamma(\nu+\frac{n}{2})\Gamma(N-\nu)}{
\Gamma(N+n+\nu)\Gamma(-\nu)}Y_{\alpha}^N({\eta})Y_{\alpha}^N({\eta}')
\end{equation}

\begin{equation}
[|\frac{a_{{\eta}'}}{a_B}{\eta}- \frac{a_B}{a_{{\eta}'}}{\eta}'|^2]^{\nu} = 
\sum_{N=0}^{N_0}\sum_{\alpha=0}^{N}
\frac{(2a)^{2\nu+n}{\pi}^{\frac{n}{2}}\Gamma(\nu+\frac{n}{2}+
\frac{1}{N_0})\Gamma(N-\nu+ \frac{1}{N_0})}{
\Gamma(N+n+\nu+ \frac{1}{N_0})\Gamma(-\nu)}Y_{\alpha}^N({\eta})Y_{\alpha}^N({\eta}')
\end{equation}
for the powers appearing in (10). Moreover, in the context of the spherical formulation 
developed in \cite{Tsoupros} the orthonormality condition for the n-dimensional spherical 
harmonics defined on $ S_n$

\begin{equation}
\int_S d^{n}{\eta}Y_{\alpha}^N(\eta)Y_{\alpha'}^{N'}(\eta) =
\delta_{NN'}\delta_{\alpha \alpha'}
\end{equation} 
is, necessarily, replaced on $ C_n$ in terms of Gegenbauer polynomials 
$ C_m^p(cos \theta)$ with their explicit dependence on the angles 
associated with the embedding $ (n+1)$-vector $ \eta$ 

$$
\int_C d^{n}{\eta}Y_{\alpha}^N(\eta)Y_{\alpha'}^{N'}(\eta) = AKa
(sin{\theta_n^0})^{m_1+l_1}C_{N-m_1}^{m_1+\frac{n-1}{2}}(cos{\theta_n^0})
C_{N'-l_1}^{l_1+\frac{n-1}{2}}(cos{\theta_n^0})\delta^{m_1l_1}\delta_{aa'}~~ + $$

$$ 
A{B}
\prod_{k=1}^{n-2}\int_0^{\pi}C_{m_k-m_{k+1}}^{m_{k+1}+\frac{n-2}{2}}(cos{\theta_{n-k}})
C_{l_k-l_{k+1}}^{l_{k+1}+\frac{n-2}{2}}(cos{\theta_{n-k}})[sin{\theta_{n-k}}]^{m_{k+1}+l_{k+1}+n-3}
d{\theta_{n-k}}\times $$

$$
(14a)~~~~~~~~~~\hspace{2.5in} 
\int_0^{2\pi}\frac{e^{i(\pm m_{n-1}\mp l_{n-1})\theta_1}}{sin{\theta_1}}
\hspace{3in}
$$
 with $ \theta_n^0$ being the angle associated with $ \partial C_n$, with 

$$
(14b)~~~~~~~~~~\hspace{2.0in} 
m_1 = Ncos{\theta_n^0}~~;~~l_1 = N'cos{\theta_n^0}
\hspace{3in}
$$
being the degrees of spherical harmonics defined on $ \partial C_n$, with

$$
(14c)~~~~~~~~~~\hspace{2.5in}
\delta_{\alpha \alpha'} = \delta^{m_1l_1}\delta_{aa'} 
\hspace{3in}
$$
and with

$$
(14d)~~\hspace{1.5in}
A = \frac{1}{(N'+\frac{n}{2}-1)(N'+\frac{n}{2}) - (N+\frac{n}{2}-1)(N+\frac{n}{2})};~~N \neq N' 
\hspace{1.5in}$$ 
and 

$$  
{B} = 2\big{[}m_1(sin{\theta_n^0})^{m_{1}+l_{1}+n-2}(cos{\theta_n^0})
C_{N-m_{1}}^{m_1+\frac{n-1}{2}}(cos{\theta_n^0})~~ - $$

$$
(14e)\hspace{1.5in}  
(sin{\theta_n^0})^{m_{1}+l_{1}+n}(2m_1+n-1)C_{N-m_{1}-1}^{m_1+\frac{n-1}{2}+1}(cos{\theta_n^0})\big{]}
C_{N'-l_{1}}^{l_1+\frac{n-1}{2}}(cos{\theta_n^0})
$$ 

The first term in the  integrand of (10) stems exclusively from the fundamental part of the 
propagator $ D_{c}^{(n)}(\eta,{\eta}')$ and is, for that matter, identical to the corresponding
term which upon integration over the volume of $ S_n$ yields the entire contribution to the 
zero-point function at $ O(\lambda^2)$ on the Euclidean de Sitter space. As, in the present
case, the integral is taken over a bounded manifold a comparison of this mathematical 
expression on $ S_4$ and $ C_4$ would be explicitly indicative of the mathematical origin of 
boundary-related contributions to divergences stemming from vacuum effects. The corresponding 
integral for the contribution to the zero-point function on $ S_n$, for that matter, is

\addtocounter{equation}{1}%
\begin{equation}
I_S = \lambda^2 B_n^4 {\int}_S d^n{\eta}d^n{\eta'}[|{\eta}-{\eta}'|^2]^{\frac{8-4n}{2}}       
\end{equation}
which, in the context of (11), yields

\begin{equation}
I_S = \lambda^2 B_n^4 
\sum_{N=0}^{\infty}\sum_{\alpha=0}^{N}
\frac{(2a)^{8-3n}{\pi}^{\frac{n}{2}}\Gamma(\frac{8-3n}{2})\Gamma(N-\frac{8-4n}{2})}{
\Gamma(N+\frac{8-2n}{2})\Gamma(\frac{4n-8}{2})}{\int}_S d^n{\eta}d^n{\eta'}
Y_{\alpha}^N({\eta})Y_{\alpha}^N({\eta}')
\end{equation}
whereupon, due to the orthonormality condition (13) and the fact that $ Y_0^0$ relates  
to the embedding radius $ a$ and to the solid angle $ \Omega_{n+1}$ for $ S_n$ through  
\cite{Bateman}

\begin{equation}
Y_{0}^{0} = \frac{1}{\sqrt{a^n \Omega_{n+1}}}
\end{equation}
it further reduces to

$$
I_S = \lambda^22^{-3n}\pi^{\frac{-3n}{2}}a^{8-3n}
\frac{[\Gamma(\frac{n}{2}-1)]^4\Gamma(4-\frac{3n}{2})}{\Gamma(4-n)}
\int_S d^{n}{\eta}
$$
where use has been made of (9) as well. It is evident that at $ n \rightarrow 4$
the pole contained in $ \Gamma(4-\frac{3n}{2})$ cancels against that in
$ \Gamma(4-n)$. Setting $ \epsilon = 4 - n$ and using the standard expansion

\begin{equation}
\Gamma(\epsilon) = \frac{1}{\epsilon} - \gamma + 
\frac{\frac{\pi^2}{6} + {\gamma}^2}{2}\epsilon + ...
\end{equation}
for $ \epsilon << 1$ the final expression at $ n \rightarrow 4$ is

\begin{equation}
I_S = \lambda^2 \frac{1}{3^32^{16}\pi^6}R^2
\end{equation}  
where use has been made of (3) and a multiplicative factor amounting to the
volume of $ S_n$ is implied. 

The result expressed in (19) amounts to a finite contribution to the gravitational effective action 
on $ S_4$. It becomes evident, for that matter, that the zero-point function has no 
divergent contribution to the gravitational component of the action on $ S_4$ to second 
order in the self-coupling $ \left[ O(\hbar^3) \right]$. This result is consistent with \cite{BrowColl} 
where, in fact, it was shown that on a general four-dimensional unbounded curved manifold vacuum 
effects due to a $ \phi^4$ self-interaction do not generate divergent contributions to 
the $ R^2$ sector of the gravitational action before the fifth order in $ \lambda$, provided that dimensional 
regularisation is used.

The same procedure applied on the same term integrated over the volume of $ C_n$ results in the 
following expression for the first term in (10)

$$
\lambda^2B_n^4 \int_C d^n{\eta}d^n{\eta'}\big{[}|\eta - \eta'|^2\big{]}^{\frac{8-4n}{2}} = $$

\begin{equation}
\lambda^2B_n^4 
\sum_{N=0}^{\infty}\sum_{\alpha=0}^{N}
\frac{(2a)^{8-3n}{\pi}^{\frac{n}{2}}\Gamma(\frac{8-3n}{2})\Gamma(N-\frac{8-4n}{2})}{
\Gamma(N+\frac{8-2n}{2})\Gamma(\frac{4n-8}{2})}
\int_C d^n{\eta}d^n{\eta'}Y_{\alpha}^N(\eta)Y_{\alpha}^N(\eta')
\end{equation}
where, again, the double integral is to be tackled on the basis of (17)

$$ 
\int_C d^n{\eta}Y_{\alpha}^N(\eta) = \sqrt{a^n \Omega_{n+1}} \int_C d^n{\eta}Y_{\alpha}^N(\eta)Y_0^0
$$
on the understanding that the orthonormality condition (13) on $ S_n$ is replaced by the condition (14) 
on $ C_n$. In this case, however, the constancy of $ Y_0^0$ enforces a 
vanishing result on the intractable second term of (14a). Rather than deal with that term 
itself a faster expedient to this result is provided directly by the reduction 
formula \cite{Tsoupros}
 
\begin{equation}
\int_C d^{n}{\eta}Y_{\alpha}^N(\eta)Y_{\alpha'}^{N'}(\eta) = Aa^2
\oint_{\partial C} d^{n-1}{\eta}[KY_{\alpha}^N(\eta)Y_{\alpha'}^{N'}(\eta) + 
2n_pY_{\alpha'}^{N'}(\eta)D_pY_{\alpha}^{N}(\eta)]
\end{equation}
which is, itself, a condition for (14) and in whose context it becomes obvious that for
$ Y_{\alpha'}^{N'} = Y_0^0$ the stated second term merely contributes a factor of $ 
-2KY_{\alpha}^N(\eta)Y_{0}^{0}$ in the integrand. Consequently,

$$ 
\int_C d^n{\eta}Y_{\alpha}^N(\eta)Y_0^0 = - aK 
\frac{(sin{\theta_n^0})^{m_1}C_{N-m_1}^{m_1+\frac{n-1}{2}}(cos{\theta_n^0})
C_{0}^{\frac{n-1}{2}}}{(\frac{n}{2}-1)\frac{n}{2}-(N+\frac{n}{2}-1)(N+\frac{n}{2})}
\delta^{m_10}\delta_{a0} ;~~N \neq 0  $$
which, through the relation \cite{Bateman}

$$
C_0^{k}(x) = 1; k > - \frac{1}{2} $$
reduces to 

$$ 
\int_C d^n{\eta}Y_{\alpha}^N(\eta)Y_0^0 = - aK 
\frac{1}{(\frac{n}{2}-1)\frac{n}{2}-(N+\frac{n}{2}-1)(N+\frac{n}{2})}
C_N^{\frac{n-1}{2}}(cos \theta_n^0) $$
so that the double integral over $ C_n$ in (20) eventually becomes

\begin{equation}
\int_C d^n{\eta}d^n{\eta'}Y_{\alpha}^N(\eta)Y_{\alpha}^N(\eta') = 
\frac{a^2K^2}{[(\frac{n}{2}-1)\frac{n}{2}-(N+\frac{n}{2}-1)(N+\frac{n}{2})]^2}
\big{[}C_N^{\frac{n-1}{2}}(cos \theta_n^0) \big{]}^2
a^n \Omega_{n+1}
\end{equation}

In the context of (20), (22), (14c) and (8) the first term in (10) assumes the
form

$$
\lambda^2B_n^4 \int_C d^n{\eta}d^n{\eta'}\big{[}|\eta - \eta'|^2\big{]}^{\frac{8-4n}{2}} =
\lambda^2 a^{10-2n} K^2 \Omega_{n+1} 
[\frac{\Gamma(\frac{n}{2}-1)}{4\pi^{\frac{n}{2}}}]^4 $$

\begin{equation}
\sum_{N=0}^{\infty}\frac{\big{[}C_N^{\frac{n-1}{2}}(cos \theta_n^0) \big{]}^2}{[(\frac{n}{2}-1)\frac{n}{2}-
(N+\frac{n}{2}-1)(N+\frac{n}{2})]^2} 
\frac{(2)^{8-3n}{\pi}^{\frac{n}{2}}\Gamma(\frac{8-3n}{2})\Gamma(N-\frac{8-4n}{2})}{
\Gamma(N+\frac{8-2n}{2})\Gamma(\frac{4n-8}{2})};~~N \neq 0
\end{equation}
whereupon setting $ \epsilon = 4-n$ and using (18) prior to taking $ \epsilon \rightarrow 0$ renders 
the first term of (10) in the form

$$
\lambda^2B_n^4 \int_C d^n{\eta}d^n{\eta'}\big{[}|\eta - \eta'|\big{]}^{4(2-n)} \rightarrow $$

\begin{equation}
\lambda^2a^2K^2\Omega_{5}2^{-13}3^{-2}\pi^{-6}\big{[}\sum_{N=1}^{\infty}[C_N^{\frac{3}{2}}(cos{\theta_4^0})]^2
\frac{(N+1)(N+2)}{N(N+3)}\big{]}\frac{1}{\epsilon} + F.T.
\end{equation}
where use has, also, been made of the standard expressions for $ \Gamma$ functions of an integer argument $ m$ and of a complex argument $ z$ respectively

$$
\Gamma(m) = (m-1)! ~~;~~z\Gamma(z) = \Gamma(z+1)$$

The series of any power of $ C_N^k(x)$ over the degree $ N$ for a fixed order $ k$ has a radius of 
convergence between $ (-1, 1)$ \cite{Bateman}. For that matter, the series featured in (24), 
is convergent and the first term of (10) displays, as a consequence, a simple pole at $ n \rightarrow 4$. 

The physical interpretation of (24) becomes obvious through the observation that the volume of the embedded $
 S_4$ relates to the solid angle $ \Omega_5$ through  

$$
\int_C d^4{\eta}+\int_{S-C} d^4{\eta} = a^4\Omega_5 $$
with

$$
\int_{S-C} d^4{\eta} = V_c\int_C d^4{\eta};~~V_c > 0 $$

%
%
%
%
%
%
%
%
%

which renders (24) exclusively in terms of volume integration over $ C_4$ 

$$
\lambda^2B_n^4 \int_C d^n{\eta}d^n{\eta'}\big{[}|\eta - \eta'|\big{]}^{4(2-n)} \rightarrow $$

$$
\lambda^2\frac{1}{a^2}K^22^{-13}3^{-2}\pi^{-6}\big{[}\sum_{N=1}^{\infty}[C_N^{\frac{3}{2}}(cos{\theta_4^0})]^2
\frac{(N+1)(N+2)}{N(N+3)}\big{]}\frac{1}{\epsilon}\int_C d^4{\eta} + F.T.
$$
allowing for a multiplicative factor which depends exclusively on the volume of $ C_4$. Use of (3) 
then finally yields at $ n \rightarrow 4$ 

$$
\lambda^2B_n^4 \int_C d^n{\eta}d^n{\eta'}\big{[}|\eta - \eta'|\big{]}^{4(2-n)} \rightarrow $$

\begin{equation}
\lambda^2\frac{1}{2^{15}3^{3}\pi^{6}}\left[ \sum_{N=1}^{\infty}[C_N^{\frac{3}{2}}(cos{\theta_4^0})]^2
\frac{(N+1)(N+2)}{N(N+3)}\right] (RK^2)\frac{1}{\epsilon}\int_C d^4{\eta} + F.T.
\end{equation}

In sharp contrast with (19) this result reveals that volume integration over $ C_n$ of the fourth power 
of the fundamental part of the propagator in (9) entails a pole at the limit of space-time dimensionality. 
The absence of any divergent contribution to the bare gravitational action on $ S_4$ at the same loop-order 
reveals that the underlying source of the pole revealed at $ n \rightarrow 4$ is $ \partial C_4$, an effect 
which is mathematically reflected in the explicit presence of the boundary's extrinsic curvature $ K$ in (25).
In \cite{Tsoupros} the case was made to the effect that the reduction of the complete orthonormality 
condition on $ S_n$ expressed by (13) to the partial orthonormality condition on $ C_n$ expressed by (14) 
is responsible for the contributions which any divergence on $ S_4$ receives due to the presence of 
$ \partial C_4$. A comparison of the calculational context of (19) with that of (25) reveals indeed that, 
in the context of dimensional regularisation, the finite contribution to the effective action on $ S_4$ 
emerges as a result of the pole cancellation relevant to the numerator-related $ \Gamma(\frac{8-3n}{2})$ 
and the denominator-related $ \Gamma(N+ \frac{8-2n}{2})$ in (16). That cancellation, however, which 
essentially stems from (13) on $ S_4$ is unattainable on $ C_4$ where that orthonormality condition 
``breaks down'' allowing, as a result, for an infinite number of $ N$-dependent terms in addition to the 
cancellation-associated $ N=0$ term. Moreover, since the replacement of (13) by (14) responsible for that effect 
is, itself, a direct consequence of the presence of the boundary $ \partial C_4$ it becomes evident that all 
boundary-related contributions to divergences already present on $ S_4$ are, necessarily, effected by the volume 
integral which, in the context of the Feynman rules, stems from vertex integration in the configuration space of 
$ C_4$ \cite{Tsoupros}.   

The second, third and fourth term in (10) involve products of powers associated
with both the fundamental and boundary part of the propagator in (9). They necessitate, for
that matter, a simultaneous use of (11) and (12) on the understanding that at the dimensional 
limit $ n \rightarrow 4$ all divergences stem exclusively from (11), a fact which is ensured by
the presence of the inverse of the upper limit $ N_0$ related to the cut-off angular momentum 
for image propagation in the arguments of the $ \Gamma$ functions in (12)  \cite{Tsoupros}.
The simultaneous substitutions through (11) and (12) are expected to result in mathematical
expressions which are substantially more involved than that of the, hitherto tackled, first term. These 
expressions necessarily relate to the residues of the resulting poles and to finite terms 
of the associated expansions and do not, for that matter, complicate the pole-structure of the 
theory. In effect, a direct application of (11), (12) and (8) on the second term of (10) results in

$$
\lambda^2B_n^4 4\int_C d^n{\eta}d^n{\eta'}|{\eta}-{\eta}'|^{3(2-n)}|\frac{a_{{\eta}'}}{a_B}{\eta}-
\frac{a_B}{a_{{\eta}'}}{\eta}'|^{2-n} = \lambda^2[\Gamma(\frac{n}{2}-1)]^4\pi^{-2n}2^{-6}\times  $$

$$
\sum_{N=0}^{\infty}\sum_{\alpha=0}^{N}
\frac{(2a)^{6-2n}{\pi}^{\frac{n}{2}}\Gamma(\frac{6-2n}{2})\Gamma(N-\frac{6-3n}{2})}{
\Gamma(N+\frac{6-n}{2})\Gamma(\frac{3n-6}{2})}
\sum_{N'=0}^{N'_0}\sum_{\alpha'=0}^{N'}
\frac{(2a)^{2}{\pi}^{\frac{n}{2}}\Gamma(1+ \frac{1}{N'_0}) \Gamma(N'-\frac{2-n}{2}
+ \frac{1}{N'_0})}{\Gamma(N'+\frac{2+n}{2}+\frac{1}{N'_0})\Gamma(\frac{n-2}{2})}\times  $$

\begin{equation}
\int_C d^n{\eta}Y_{\alpha}^N({\eta})Y_{\alpha'}^{N'}({\eta})\int_C d^n{\eta'}Y_{\alpha}^N({\eta'})Y_{\alpha'}^{N'}({\eta'})
\end{equation} 
Since each of the integrals over the product of spherical harmonics is directly expressed
by (14) it is

$$ 
\int_C d^n{\eta}Y_{\alpha}^N({\eta})Y_{\alpha'}^{N'}({\eta})\int_C d^n{\eta'}Y_{\alpha}^N({\eta'})Y_{\alpha'}^{N'}({\eta'}) = $$

$$
\frac{1}{[(N'+\frac{n}{2}-1)(N'+\frac{n}{2}) - (N+\frac{n}{2}-1)(N+\frac{n}{2})]^2} \times
$$

\begin{equation}
\left[a^2K^2F^2(n)\delta_{\alpha \alpha'}  + 2aK[F(n)B(n)H(n)]\delta_{\alpha \alpha'} + [B(n)H(n)]^2\right]
\end{equation} 
with the condition $ N \neq N' $ and with  

$$
F(n) = (sin{\theta_n^0})^{2m_1}C_{N-m_1}^{m_1+\frac{n-1}{2}}(cos{\theta_n^0})
C_{N'-m_1}^{m_1+\frac{n-1}{2}}(cos{\theta_n^0}) ~~;~~m_1 = Ncos{\theta_n^0},~~l_1 = N'cos{\theta_n^0}$$
and

$$
H(n) = \prod_{k=1}^{n-2}\int_0^{\pi}C_{m_k-m_{k+1}}^{m_{k+1}+\frac{n-2}{2}}(cos{\theta_{n-k}})
C_{l_k-l_{k+1}}^{l_{k+1}+\frac{n-2}{2}}(cos{\theta_{n-k}})[sin{\theta_{n-k}}]^{m_{k+1}+l_{k+1}+n-3}
d{\theta_{n-k}}\times $$

$$
\int_0^{2\pi}\frac{e^{i(\pm m_{n-1}\mp l_{n-1})\theta_1}}{sin{\theta_1}} $$
 
Substituting that expression in (26) and taking the limit $ 4-n = \epsilon \rightarrow 0$ 
results in a simple pole contained in $ \Gamma(\frac{6-2n}{2})$

$$
\lambda^2B_n^4 4\int_C d^n{\eta}d^n{\eta'}|{\eta}-{\eta}'|^{3(2-n)}|\frac{a_{{\eta}'}}{a_B}{\eta}-
\frac{a_B}{a_{{\eta}'}}{\eta}'|^{2-n}  \rightarrow $$

$$
\lambda^2\frac{(-1)}{2^{7}\pi^{4}}\sum_{N=0}^{\infty}\sum_{N'=0}^{N'_0}
\frac{\Gamma(N+3)}{\Gamma(N+1)}\frac{\Gamma(1+\frac{1}{N'_0})\Gamma(N'+1+
\frac{1}{N'_0})}{\Gamma(N'+3+\frac{1}{N'_0})}\frac{1}{[{N'}^2-N^2+3(N'-N)]^2}\times$$

$$
\left[a^2K^2F^2(4)\sum_{\alpha=0}^{N}\sum_{\alpha'=0}^{N'}\delta_{\alpha \alpha'}
+ 2aK[F(4)B(4)H(4)]\sum_{\alpha=0}^{N}\sum_{\alpha'=0}^{N'}\delta_{\alpha \alpha'}
+ [B(4)H(4)]^2\right]\frac{1}{\epsilon} + F.T.
$$
which, in view of the double summation over the Kronecker symbol and the condition $ N \neq N'$, amounts to

$$
\lambda^2B_n^4 4\int_C d^n{\eta}d^n{\eta'}|{\eta}-{\eta}'|^{3(2-n)}|\frac{a_{{\eta}'}}{a_B}{\eta}-
\frac{a_B}{a_{{\eta}'}}{\eta}'|^{2-n}  \rightarrow $$

$$
\lambda^2\frac{(-1)}{2^{7}\pi^{4}}\sum_{N=0}^{\infty}\sum_{N'=0}^{N'_0}
\frac{\Gamma(N+3)}{\Gamma(N+1)}\frac{\Gamma(1+\frac{1}{N'_0})\Gamma(N'+1+
\frac{1}{N'_0})}{\Gamma(N'+3+\frac{1}{N'_0})}\frac{1}{[{N'}^2-N^2+3(N'-N)]^2}\times$$

$$
\left[a^2K^2F^2(4)N' + 2aK[F(4)B(4)H(4)]N' + [B(4)H(4)]^2\right]\frac{1}{\epsilon} + F.T.
$$
whereupon through use of 

$$
\int_C d^4{\eta} = a^4\Omega_5^{(C)} $$
in the first and third term of the geometry-dependent multiplicative factor and

$$
\oint_{\partial C} d^{3}{\eta} = \big{(}asin(\theta_4^0)\big{)}^3\Omega_4 $$
in the second term of that factor and through (3) for $ n = 4$ the result for
the second term in question is

$$
\lambda^2B_n^4 4\int_C d^n{\eta}d^n{\eta'}|{\eta}-{\eta}'|^{3(2-n)}|\frac{a_{{\eta}'}}{a_B}{\eta}-
\frac{a_B}{a_{{\eta}'}}{\eta}'|^{2-n}  \rightarrow $$

$$
\lambda^2\frac{(-1)}{2^{7}\pi^{4}}\sum_{N=0}^{\infty}\sum_{N'=0}^{N'_0}
\frac{\Gamma(1+\frac{1}{N'_0})\Gamma(N'+1+
\frac{1}{N'_0})}{\Gamma(N'+3+\frac{1}{N'_0})}\frac{(N+1)(N+2)}{[{N'}^2-N^2+3(N'-N)]^2}\times$$

$$
\big{[} \frac{1}{3}\frac{1}{2^2}F^2(4)N' (RK^2) (\Omega_5^{(C)})^{-1}\int_Cd^4{\eta} + \frac{1}{3}\frac{1}{2} 
[F(4)B(4)H(4)]N' (RK) (\Omega_4)^{-1}\big{(}sin(\theta_4^0)\big{)}^{-3} \oint_{\partial C} d^{3}{\eta} +  $$

\begin{equation}
\frac{1}{3^2}\frac{1}{2^4} 
[B(4)H(4)]^2 R^2 (\Omega_5^{(C)})^{-1}\int_Cd^4{\eta} \big{]} \frac{1}{\epsilon} + F.T.
\end{equation}

The infinite series over $ N$ is obviously convergent and (28) manifests, as a result, a 
simple pole at the limit of space-time dimensionality. 

The calculational procedure for the third, fourth and fifth term in (10) is identical to that for the second. The result 
for the third term is  

$$
\lambda^2B_n^4 6\int_C d^n{\eta}d^n{\eta'}|{\eta}-{\eta}'|^{2(2-n)}|\frac{a_{{\eta}'}}{a_B}{\eta}-
\frac{a_B}{a_{{\eta}'}}{\eta}'|^{2(2-n)}  \rightarrow $$

$$
\lambda^2\frac{1}{\pi^{4}}\sum_{N=0}^{\infty}\sum_{N'=0}^{N'_0}
\frac{\Gamma(\frac{1}{N'_0})}{[{N'}^2-N^2+3(N'-N)]^2}\times$$

$$
\big{[} \frac{1}{2^8}F^2(4)N' (RK^2) (\Omega_5^{(C)})^{-1}\int_Cd^4{\eta} + \frac{1}{2^7} 
[F(4)B(4)H(4)]N' (RK) (\Omega_4)^{-1}\big{(}sin(\theta_4^0)\big{)}^{-3} \oint_{\partial C} d^{3}{\eta} +  $$

\begin{equation}
\frac{1}{3}\frac{1}{2^{10}} 
[B(4)H(4)]^2 R^2 (\Omega_5^{(C)})^{-1}\int_Cd^4{\eta} \big{]} \frac{1}{\epsilon} + F.T.
\end{equation}
whereas the fourth and, as expected, fifth term each yield a finite result. Attention is 
invited at this point to the fact that the present calculation relates exclusively to the three-point
function without recourse to the possibility of an overlapping divergence stemming from the
self-coupling $ \lambda_0$ in fig.(1d). Such a divergence would, certainly, yield an additional 
contribution. The exploration of such a possibility is an issue which relates directly to the 
renormalisation of that self-coupling and constitutes the objective of the direction in which the present 
technique is to be advanced. The objective of the present calculation, however, is the contribution which 
the zero-point function and the effective action receive exclusively due to vaccuum effects at the 
specified loop-order. Allowing for finite contributions the stated entire contribution of the zero-point 
function to the bare gravitational action at third loop-order is the additive result of (25), (28) and (29).
Using \cite{Bateman} 

$$
\Omega_n = \frac{2\pi^{\frac{n}{2}}}{\Gamma(\frac{n}{2})}
$$
the stated contribution is

$$
I_c = 
\lambda^2\frac{1}{\pi^{4}}\frac{1}{2^8}\sum_{N=0}^{\infty}\big{[}
\frac{1}{\pi^{2}}\frac{1}{3^3}\frac{1}{2^7}\frac{(N+2)(N+3)}{(N+1)(N+4)} 
\big{(}C_{N+1}^{\frac{3}{2}}(cos{\theta_4^0})\big{)}^2 + $$

$$
\sum_{N'=0}^{N'_0}\frac{1}{[{N'}^2-N^2+3(N'-N)]^2}[\Gamma(\frac{1}{N'_0}) -
\frac{1}{3}\frac{1}{2}\frac{\Gamma(1+\frac{1}{N'_0})\Gamma(N'+1+\frac{1}{N'_0})}{\Gamma(N'+3+\frac{1}{N'_0})}(N+1)(N+2)] \times $$

$$
3\frac{1}{2^3}\frac{1}{\pi^2}
F^2N'\big{]}\frac{RK^2}{\epsilon}\int_Cd^4{\eta} + $$

$$
\lambda^2\frac{1}{\pi^{6}}\frac{1}{2^{13}}\sum_{N=0}^{\infty}\sum_{N'=0}^{N'_0}
\frac{1}{[{N'}^2-N^2+3(N'-N)]^2} \times $$

$$
\left[ \Gamma(\frac{1}{N'_0}) - 
\frac{1}{3}\frac{1}{2}\frac{\Gamma(1+\frac{1}{N'_0})\Gamma(N'+1+\frac{1}{N'_0})}{\Gamma(N'+3+\frac{1}{N'_0})}(N+1)(N+2) \right](BH)^2
\frac{R^2}{\epsilon}\int_Cd^4{\eta} +  $$

$$
\lambda^2\frac{1}{\pi^{6}}\frac{1}{2^{8}}\sum_{N=0}^{\infty}\sum_{N'=0}^{N'_0}
\frac{1}{[{N'}^2-N^2+3(N'-N)]^2} \times $$

$$
\left[ \Gamma(\frac{1}{N'_0}) - 
\frac{1}{3}\frac{1}{2}\frac{\Gamma(1+\frac{1}{N'_0})\Gamma(N'+1+\frac{1}{N'_0})}{\Gamma(N'+3+\frac{1}{N'_0})}(N+1)(N+2) \right]\times $$

\begin{equation}
(FBH)N'\big{(}sin(\theta_4^0)\big{)}^{-3}
\frac{RK}{\epsilon}\oint_{\partial C}d^3{\eta} 
\end{equation}
with the condition $ N \neq N'$ and with $ n=4$ in all n-dependent quantities.

{\bf III. Discussion}

As expected \cite{George}, \cite{Tsoupros} radiative contributions on $ C_4$ are substantially more involved than 
the corresponding contributions on $ S_4$. The complicated mathematical expression for the radiative contributions 
in question are the necessary consequence of both the presence of $ \partial C_4$ and the much smaller symmetry 
underlying the geometry of $ C_4$. 
However, all ultra-violet divergences at third loop-order are expressed in terms of simple poles. The 
result expressed by (30) reveals the stated contribution of the zero-point function as the additive result of three 
ultra-violet divergent contributions associated with the generation of three distinct sectors in the bare and, 
through renormalisation, effective gravitational action. In addition to the generation of the volume divergence 
proportional to $ RK^2$ there is also a volume divergence proportional to $ R^2$ as well as a surface divergence 
proportional to $ RK$. Since the one-loop contribution to the two point function, expressed by fig.(1a) and fig.(1b),
is finite in the context of dimensional regularisation the associated three sectors constitute quantum corrections to
the Einsten-Hilbert action generated by vacuum effects of the self-interacting scalar field at loop-orders no smaller 
than the third. In the context of the comparison with $ S_4$ the volume divergence proportional to $ RK^2$ has been 
exposed as the exclusive result of the presence of the boundary and, for that matter, of the cummulative effect of 
reflection off the boundary of signals signifying propagation on $ S_4$. The presence of such a divergence in the 
theory is indicative of the contributions which any divergence arising from the coincidence of space-time points 
receives from the boundary on a general bounded manifold in addition to the contributions effected by the 
background geometry. With respect to the latter divergences proportional to the square of the Ricci scalar are
expected on a general curved manifold \cite{Birrel}. However, the generation of that volume divergence on $ C_4$ 
already at second order in the self-coupling $ \lambda$ contrasts, again, strongly with the result reported in 
\cite{BrowColl} regarding the absence of any divergent contributions to the $ R^2$ sector of the gravitational 
action on a general curved space-time without boundary before the fifth order in $ \lambda$ in the context of 
dimensional regularisation. The generation of a volume ultra-violet divergence substantially earlier in the 
perturbative expansion and, for that matter, at significantly lower energy scales is again a boundary-related 
effect. It becomes evident, for that matter, that the divergences inherent in the zero-point function evaluated 
at three-loop level receive non-trivial contributions from the boundary $ \partial C_4$ which, effectively, 
dissociates the dynamical behaviour of the quantised scalar field on $ C_4$ from that on $ S_4$. In the same 
respect the emergence of the additional surface term in the bare gravitational action as a result of 
the remaining surface $ RK$-related ultra-violet divergence constitutes a purely quantum effect generated by
vacuum processes in the volume of $ C_4$. As is the case with the $ RK^2$-related sector this is also a 
qualitatively new result effected at third loop-order by the presence of the boundary. The $ RK$-related sector
in the bare gravitational action signifies, in fact, the first surface counterterm generated by vacuum effects 
on a general manifold with boundary. The absence of surface counterterms on a general manifold with boundary 
has been reported at one-loop level \cite{Od}, \cite{Odintsov}. However, the existence of such counterterms
has been anticipated at higher loop-orders on theoretical grounds \cite{George} and the potential for their
generation has been confirmed in the context of the spherical formulation for diagramatic evaluations on a 
manifold with boundary \cite{Tsoupros}. The emergence of this new sector in the bare gravitational action 
confirms such expectations.    

Two more mathematical features of the radiative contributions expressed by (30) deserve attention in their own merit.
In addition to the emergence of boundary-related ultra-violet divergences the theory is characterised by the absence 
of infra-red divergences on $ C_4$. In light of (14) this result is, evidently, relevant to all orders \cite{Tsoupros}. 
Moreover, on the evidence of (30), all radiative contributions manifest an explicit dependence on the cut-off angular 
momentum for image propagation $ N'_0$. Such dependence is expected on a general compact bounded manifold and, 
necessarily, reflects the effect which the geodesic distance between points on the boundary has on the dynamical 
behaviour of the quantised fields. 

The issue which arises as a consequence of the preceding considerations is the degree to which the results
relevant to (30) can be further generalised on any curved bounded manifold. The merit of the spherical
formulation which has eventuated in (30) is the expression of all Green functions of the theory in terms of 
the constant scalar curvature $ R$, a result which has both formal and perturbative significance. In effect, 
the additional volume divergences - proportional to the Ricci and Riemann tensor-related contractions 
$ R_{\mu\nu}R^{\mu\nu}$ and 
$ R_{\alpha\beta\gamma\delta}R^{\alpha\beta\gamma\delta}$ respectively - which are expected on a general curved 
manifold are precluded on both $ S_4$ and $ C_4$. However, the significance of divergences such as those contained in 
(30) is, as stated, general. Their presence in this theory in the context of the demand for dimensional consistency 
renders them relevant to a general curved bounded manifold. Consequently, the stated K-dependent sectors are formally 
expected in the bare action on any curved bounded manifold and their presence in this theory is indicative of the 
non-trivial effect which the topology of a general bounded manifold has on the cut-off scale behaviour of quantised 
matter. 

A relevant objection which might be raised at this point is associated with an essential aspect of the mathematical 
approach through which these results have been obtained. Specifically, despite its stated geometry-related 
generality such a conclusion would appear to still leave open the question of the regulating procedure. As stated, 
all ultraviolet divergences have been regulated through an analytical extension of the space-time dimensionality to 
an arbitrary value $ n$. That approach preserves the maximal symmetry which underlies the spherical formulation on an 
arbitrary $ S_n$ and constitutes, for that matter, the natural generalisation of $ S_4$ within the framework of 
dimensional regularisation. Such an extension, however, is by no means unique. The procedure of dimensional 
regularisation is, in fact, known to manifest a certain degree of ambiguity in curved space-time due to its inherent 
freedom in generalising the associated dimensionality within frameworks of different topologies \cite{Hawking},
\cite{DowkerCr}, \cite{Toms}. In the present context, for example, equally valid a procedure would be the formal 
extension of $ S_4$, the covering manifold of the physical space-time $ C_4$, in dimensions of flat topology. 
Such an approach would eventuate in a regulating scheme predicated on $ S_4 \times R_{n-4}$ which, obviously, 
does not sustain the $ SO(n+1)$ de Sitter invariance manifest in (1).  

The ambiguity resulting from the stated freedom of choice stems, essentially, from the effect which the presence
of a non-trivial topology potentially has on the divergent structure of the theory. Such an effect is a peculiarity 
of the quantised theory since on the basis of the equivalence principle only local curvature terms would be expected 
to be of relevance to the renormalisation program. The non-trivial effect which the, global, topology potentially 
manifests has called into question the validity of the equivalence principle in the context of second quantisation 
\cite{Toms}, \cite{Tom}, \cite{To}, \cite{DrummHathr}, \cite{BirrFord}. In fact, as has already been stressed, the 
results obtained for the zero-point function in the context of the present calculation are indicative of such a 
non-trivial impact on the renormalisation program. In the context of dimensional regularisation such a
topology-related impact may still persist at the limit of space-time dimensionality. In effect, both the pole 
structure and the resulting value for a closed-loop graph would, in general, be different for these different 
extensions in $ n$ dimensions. Such a difference, however, is not expected to affect the physical content of the 
theory. Topologically different extensions in the context of dimensional regularisation would affect, for that matter, 
neither the physical amplitudes manifest in the Green functions nor the renormalisation group behaviour of the theory.
The perturbative evaluation of the theory's Green functions is expected to remain unaffected up to, at most, a 
finite renormalisation.   

In view of these considerations a dimensional extension of $ S_4$ to 
$ S_4 \times R_{n-4}$ as opposed to $ S_n$ 
prior to the analytical extension of $ n$ in the context of dimensional 
regularisation may potentially generate a 
result for the three-loop contribution to the zero-point function which differs 
from that of (30) in pole-structure 
and finite part. However, the overall contribution to the zero-point function 
at three-loop level will remain 
unaffected by these differences. It is stressed that the result in (30) 
represents the overall $ O(\hbar^3)$ contribution of the zero-point function
to the bare gravitational action. Such a contribution does not include 
possible other contributions to the zero-point function stemming from the 
scalar action. The emergence of a divergent structure 
different from that of (30) if 
$ n \rightarrow 4$ is taken in the context of $ S_4 \times R_{n-4}$ would be 
expected to cancel against the
different structure which would, likewise, emerge for the overlapping divergence 
stemming from the bare self-interaction vertex of fig.(2d) in a manner which 
would eventuate in the same pole-structure for the zero-point 
function at three-loop level as that which would be obtained were 
$ n \rightarrow 4$ to be taken in the context of $ S_n$. As a result, at 
three-loop level the gravitational effective action on $ C_4$ would be 
expected to remain unaffected by that difference in the regulating 
procedure up to a finite renormalisation. These remarks will be further 
elucidated in the context of a concrete renormalisation procedure which 
will constitute the natural extension of the results attained herein. 
Notwithstanding renormalisation, however, it might be worth conjecturing
the identity of the two regulating procedures themselves in the resulting 
pole-structures respectively. At one-loop level
the zeta-function evaluation of the effective action on a general manifold 
has been shown to be in agreement with the result generated by the extension 
of the physical dimensionality in topologically flat dimensions but not with 
that effected by extensions to other non-trivial topologies in the context of 
dimensional regularisation \cite{Hawking}. However, in the concrete case of the 
Euclidean de Sitter $ S_4$ the zeta-function evaluation of the gravitational 
effective action yields a result which is in agreement with that of dimensional regularisation 
in the context of $ S_n$. It would appear, in effect, that - 
at least at one-loop level - the two regulating procedures predicated on 
$ S_n$ and $ S_4 \times R_{n-4}$ result in the same pole-structure. An 
additional argument in favour of this identity may be elicited by the absence 
of nonlocal, curvature-independent, divergences in the topologically 
non-trivial context of $ S_1 \times R_3$ \cite{BirrFord}, \cite{Tom}, 
\cite{To}.                  

An additional ramification which the issue of the regulating procedure has 
relates directly to the coupling between the scalar field and the background 
geometry. Specifically, in conformity with dimensional regularisation, that 
coupling in the conformally invariant action (1) corresponds to the 
n-dimensional value  $ \xi (n) = \frac{1}{4}\frac{n-2}{n-1}$ rather than that 
of $ \xi = \frac{1}{6}$ which relates to the improved stress tensor in four 
dimensions. Such an approach naturally distinguishes the three-loop contribution 
to the zero-point function expressed by (30) from that effected by fixing 
$ \xi$ to its conformal value in four dimensions. The renormalised theory,
however, with its inherent one-parameter ambiguity renders the latter approach 
inconsistent. In fact, the immediate consequence of the theory's invariance 
under the renormalisation group is the scale-dependence of the theory's 
physical parameters. In effect, consistency with the renormalisation group 
invariance requires that all couplings in the theory be unspecified at the
inception of the renormalisation program and, for that matter, in the 
semiclassical action. The extension of the coupling between the scalar field 
and gravity to its conformal value in $ n$-dimensions inherent in (1) ensures 
that requirement in the context of dimensional regularisation while preserving
the classical conformal invariance in that arbitrary dimensionality. Radiative
effects can, of course, still be computed with $ \xi = \frac{1}{6}$ from the 
outset. Such an approach, however, trivialises the renormalisation group 
behaviour of the theory unless that classical value for $ \xi$ also 
corresponds to a fixed point for the renormalised theory, a situation
which amounts, essentially, to a tautology considering that the evaluation of 
any possible fixed point necessitates the solution of the renormalisation group 
equations. Again, these remarks will be further elucidated in the context of a 
concrete calculation which should render the renormalisation group behaviour of 
the theory manifest.

Pursuant to the question of generality it is also worth mentioning that results in 
perturbative renormalisation relevant to a massless scalar field conformally coupled to
the background geometry of $ C_4$ at semi-classical level with a $ \phi^4$ self-interaction are 
of immediate importance to quantum cosmology in view of the description which the ground state of 
the cosmological wavefunction admits in terms of constant-curvature gravitational instantons in 
imaginary time. The results hitherto attained on $ C_4$ up to third loop-order and the stated 
degree of generality which they, qualitatively, admit on any curved bounded manifold call into 
question the issue of renormalisation at that loop-order. It is evident from (30) that the 
ultra-violet divergences inherent in the zero-point function necessitate, at that order, the 
simultaneous introduction of both volume and surface counterterms. This result indicates that, 
in conformity with theoretical expectation \cite{George}, \cite{Tsoupros}, renormalisation in 
the volume and surface effective action is accomplished simultaneously. Such a situation 
shall be explicitly confirmed in the context of a concrete renormalisation program which, as 
stated, shall also advance the remarks made above as to the relation between the present results 
and the regulating procedure. The task of renormalising the theory past one-loop level 
is rather arduous in view of the complicated associated expressions such as that in 
(30). However, work on these lines is in progress.

\end{document}